\documentclass[pra,twocolumn,groupedaddress,showpacs]{revtex4}
\begin{document}
\newcommand{\beq}{\begin{equation}}
\newcommand{\eeq}{\end{equation}}
% You should use BibTeX and revtex.bst for references
\bibliographystyle{apsrev}

\title{Decoherence, Time Scales and Pointer States}

\author{Tabish Qureshi}
\email[Email: ]{tabish@ctp-jamia.res.in}
\affiliation{Centre for Theoretical Physics, Jamia Millia Islamia, New Delhi-110025,
INDIA.}

%%\date{\today}

\begin{abstract}

Certain issues regarding the time-scales over
which environment-induced decoherence occurs, and the nature of emergent 
pointer states, are discussed. A model system, namely, a Stern-Gerlach setup
coupled to a quantum mechanical ``heat-bath" is studied.
The emergent pointer states for this system are obtained, which are 
different from those discussed in the literature. It is pointed out that
this difference is due to some confusion regarding the decoherence 
time-scale, which is clarified here.

\end{abstract}


\maketitle

\section{Introduction}

The emergence of classical world from quantum mechanics is a subject
which is poorly understood. While the theories decribing the
classical and quantum objects are well established, the border between
the two still remains hazy for many. There have been many attempts
towards resolving this issue. Most of them either modify quantum mechanics,
e.g., the so called GRW proposal \cite{grw}, or introduce certain hidden
variables, e.g., the deBroglie-Bohm approach. 

One approach which tries to tackle the problem using only conventional
quantum mechanics forms the basis of the decoherence program, initiated
by Zeh \cite{zeh}, and later extensively developed by Zurek and collaborators
\cite{zurekrmp}.  
Decoherence approach is based on the observation that
macroscopic objects can never be considered approximately isolated. They
are always interacting weakly with various external degrees of freedom,
collectively referred to as the environment, which are normally
ignored. The interaction with the environment leads to, argues Zeh, a
dephasing of the system. This leads to certain phase relations between the
macroscopically separated parts of the wave-function, getting lost. This
information is not really lost, but leaks away almost irreversibly into
the infinitely large number of degrees of freedom of the environment. The
result is that the initial pure state density matrix of the system now
{\em appears to be} a mixed state density matrix, if one decides not
to look at the environment variables, and focuses attention only on the
system of interest. It should be noted that while the time evolution
of the system plus environment is completely unitary, the system by
itself appears to undergo a non-unitary change. This process leads to
the appearance of a classically interpretable probability distribution,
and is believed to describe the emergence of the classical world.

A prototype model that one studies in this context is that of a free particle 
coupled to a ``heat-bath" of an infinite number of independent harmonic 
oscillators\cite{feynman,dekker,legget}. This model becomes tractable
in the so-called Markovian approximation. In this limit the time evolution
of the density matrix is governed by a {\em master equation}. Zurek
did an approximate study of this master equation and showed that quantum
coherence between two pieces of the wave-function, spatially separated by
a distance $\Delta x$, is destroyed over a time-scale given by
\beq
\tau_D \approx \gamma^{-1} {\hbar^2\over 2mk_BT(\Delta x)^2},
\eeq
where $\gamma^{-1}$ is the relaxation time of the system, $k_BT$ is the thermal
energy of the heat-bath, assumed to be in equilibrium, and $m$ is the mass
of the particle. As one can see, for a particle of mass 1g, interacting with
a heat-bath at 300K, the quantum coherence over a distance of 1cm is
destroyed in a time which is $10^{-40}$ times the relaxation time.

Zurek also concluded from this analysis that the environment selects
position as a preferred basis in which the density matrix of the particle
becomes diagonal. The states constituting this basis, selected by environment 
induced decoherence,
have come to be known as {\em preferred states} or {\em pointer states}.
The nature of pointer states has also been a subject of much research attention
\cite{sieve,gogolin}.

Recently, related models have been studied in a more
rigorous manner to understand the detailed working of
decoherence\cite{rg,dk,anu1,anu}. Interestingly, the conclusions of
these studies are at variance with Zurek's approximate calculation.

The purpose of this investigation is to have a fresh look at one model
and get a clear picture regarding the decoherence time and pointer states.

\section{What to expect?}

Let us first discuss what one should expect as a result of decoherence,
if one believes that it leads to the emergence of classicality. We should
of course expect that there will be no Schr\"odinger's cats hanging around,
that is, there should be no superpositions of spatially separated states.
Zurek's analysis achieves that as he shows that quantum coherence over
macroscopic distances dies out very very fast. But that alone doesn't rule out
the existence of superpositions of macrosopically distinct momentum.
Classically we should also anticipate that quantum coherence will be
destroyed over macroscopic distances {\em in the momentum space too}.
Then classical particles will have {\em macroscopically} well-defined
position, as well as momentum, and their dynamics could be characterized
by {\em paths}.
Infact, macroscopic superpositions of no classical observable should
exist. Such should be the nature of the pointer states that
emerge.  In this light, Zurek's claim of position emerging as the preferred
basis\cite{zurek} and that of some other works\cite{rg,dk,anu1} where momentum
seems to emerge as pointer states, do not seem to satisfy the criterion of
``classicality". On the other hand, in the work of Zurek and Paz\cite{paz}
and also of Venugopalan \cite{anu} coherent states seem to emerge as
pointer states for a harmonic oscillator, which nicely satisfy the
``classicality" criterion.

We should also expect that full classical dynamics will emerge from 
decoherence, which means that the effect of interaction with the environment
will have negligibly small influence on the dynamics. It is worth pointing
out that the same models have been in use for studying quantum dissipation
\cite{dekker,legget}, where one studies the influence of a heat-bath on
the dynamics of a quantum system. Since we are assuming that our system
of interest is classically an isolated system, no dissipative effect
should be observable even if decoherence is strong. This seems like a
stringent condition, but it will turn out that this is not so.

Another thing one might expect is that
decoherence would be faster if the interaction with the environment is
stronger.
Zurek's expression for decoherence
time $\tau_D \approx \gamma^{-1} {\hbar^2\over 2mk_BT(\Delta x)^2}$ seems to
be in agreement with this line of thought. The decoherence time here is
inversely proportional to $\gamma$, which is also a measure of the strength
of interaction with the environment.

\section{The Stern-Gerlach Setup}

In the following, we look at a model, which has already been studied
before, and try to understand how decoherence works.

Consider a particle of mass $m$ and spin ${1\over 2}$, moving along the
y-axis, and passing through a magnetic field along the z-axis, which is
inhomogeneous along the x-axis. This model typically describes a 
Stern-Gerlach experiment, with a Hamiltonian 
\beq
	H_0 = {p^2\over 2m} + \epsilon x\sigma_z,
\eeq
where $x$ and $p$ denote the position and momentum operators of the particle,
and $\sigma_z$ is the Pauli spin matrix, denoting the z-component of the
spin operator, multiplied by 2.

In addition, the particle is assumed to interact with a set of infinite
number of harmonic oscillators, through the interaction
\beq
H_I = \sum_j {P_j^2\over 2m_j} 
    + {1\over 2}m_jw_j^2\left(X_j-{g_jx\over m_j\omega_j^2}\right)^2,
\label{interaction}
\eeq
which is just the Hamiltonian for a set of independent but {\em shifted}
harmonic oscillators, with the shift proportional to the position of the
particle. This in effect, represents a coordinate-coordinate coupling of the
particle with the oscillators, with $g_j$ as coupling constants. 
The aim is to study the reduced density matrix of the system by tracing
over the oscillator degrees of freedom. The off-diagonal matrix elements
represent presence of quantum coherence in the system.  

It turns out that in order to analyze the influence of the environment
consisting of harmonic oscillators on the dynamics of the particle, one
does not need individual knowledge of $g_j$s, $m_j$s and $\omega_j$s. One
only requires them in a particular combination as they appear in the
so-called spectral density function :

\begin{equation} 
J(\omega) = \sum_{j=0}^{\infty} {g_j^2\over m_j\omega_j} \delta(\omega-\omega_j) .
\end{equation} 
In order to describe the dissipative behavior that goes to the correct
classical limit, one assumes a continuous spectrum
of frequencies in the bath, with  the spectral density taking the ``Ohmic"
form where $J(\omega)$ is linear in $\omega$ for small frequencies:
\begin{equation} 
J(\omega) = \gamma\omega e^{-\omega/\omega_c} ,
\end{equation} 
where $\omega_c$ is a large cut-off frequency.

Clearly, $\gamma$ is now a measure of the strength of the coupling between
the particle and the environment. All the $g_j$s going to zero would
amount to $\gamma$ going to zero, which would represent a decoupled system
and heat-bath.

This model has been studied before by the path-integral technique \cite{rg}
and the master equation technique \cite{dk}. In both the works, the conclusion
is that the off-diagonal elements (in the spin-space) of the density matrix
decay over a time-scale
\beq
	t_s = \left({3\hbar^2m\gamma\over 2\epsilon^2k_BT}\right)^{1/3}.
\eeq
It is also concluded
that the spin-diagonal components of the density matrix become diagonal
in the {\em momentum space} over a decoherence time-scale
\beq
	\tau_D = {m\gamma\over 2k_BT(\Delta p)^2}, \label{rgtime}
\eeq
where $\Delta p$ is the spread in the momentum space.

Let us look at the the reduced density matrix for the particle more
carefully. For simplicity, let us assume that the particle started as
a Gaussian wave-packet of width $\sigma$, with mean momentum zero, meaning
the particle has no motion along x-axis when it travels along y-axis. The
spin diagonal elements of the reduced density matrix can be written as
 \cite{dk}
\begin{eqnarray}
\rho_\pm(R,r,t) &=& 2\sqrt{\pi\over M(\tau)} \exp\left\{-\left({
e^{-2\tau}\over 4\sigma^2} + {D(1-e^{-2\tau})\over 8\hbar^2\gamma}\right)r^2
\right.\nonumber\\
&\mp&{i\epsilon\over\hbar\gamma}(1-e^{-\tau})r\nonumber\\
&-&{1\over M(\tau)}\left(R \pm {\epsilon\over m\gamma^2}(1-e^{-\tau}-\tau)\right.\nonumber\\
&-&\left.\left.{i\hbar r\over 2\sigma^2m\gamma}e^{-\tau}(1-e^{-\tau})
+ {iDr\over 4m\gamma^2\hbar}(1-e^{-\tau})^2 \right)^2\right\}, \nonumber\\
\end{eqnarray}
where $\tau = \gamma t$, $D=8m\gamma k_BT$, $r=x-x'$, $R={x+x'\over 2}$ and
\beq
M(\tau) = \sigma^2 + {\hbar^2(1-e^{-\tau})^2\over\sigma^2m^2\gamma^2}
+ {D\over 2m^2\gamma^3}(2\tau-3+4e^{-\tau}-e^{-2\tau}).
\eeq

Now, this is an exact solution of the master equation, and needs to be
looked at in some appropriate limit which satisfy the criteria we spelt
out in section II. In order that dissipative effects do not show up
in the dynamics, the relaxation time $\gamma^{-1}$ must be much much
larger than any other time of interest, i.e., $\gamma t \ll 1$. We
could, for example, easily consider $\gamma^{-1}$ to be 100 years, and
then look at the dynamics for any long time meaningful in the lab.  So,
for $\gamma t \ll 1$, the expression for the spin-diagonal components of
the density matrix takes the approximate form
\begin{eqnarray}
\lefteqn{\rho_\pm(R,r,t) \approx}\nonumber\\
&&  2\sqrt{\pi\over \sigma^2} \exp\left\{-{2m\gamma k_BT(x-x')^2\over \hbar}t)
\right.\nonumber\\
 &&\times \exp\left(\mp{i\epsilon xt\over\hbar}-{1\over 2\sigma^2}
   (x\mp {\epsilon t^2\over 2m}-{i\hbar rt\over 2\sigma^2m}
   +{iDrt^2\over 4m\hbar})^2\right)\nonumber\\
 &&\times \exp\left(\pm{i\epsilon x't\over\hbar}-{1\over 2\sigma^2}
   (x'\mp {\epsilon t^2\over 2m}-{i\hbar rt\over 2\sigma^2m}
   +{iDrt^2\over 4m\hbar})^2\right). \nonumber\\
\end{eqnarray}
One can see that there is an exponential decay term which destroys parts
with finite $(x-x')^2$, and leads to decoherence. And this happens over
a time-scale
\beq
\tau_D = {\hbar^2\over 2m\gamma k_BT(x-x')^2}.
\eeq
One can see that Zurek's approximate treatment is in agreement with this,
and so is the path-integral calculation of Banerjee and Ghosh \cite{rg}.
The imaginary terms in the arguments of the two Gaussians depend on $(x-x')$,
and hence become inconsequential because of fast destruction of terms
involving $(x-x')$. If one ignores them, the density matrix, rather the spin
diagonal components of it, look like
\begin{eqnarray}
\rho_\pm(R,r,t) &\approx&
 2\sqrt{\pi\over \sigma^2} \exp(-{2m\gamma k_BT(x-x')^2\over \hbar}t)\nonumber\\
 &&\times \exp\left(\mp{i\epsilon xt\over\hbar}-{1\over 2\sigma^2}
   (x\mp {\epsilon t^2\over 2m})^2\right)\nonumber\\
 &&\times \exp\left(\pm{i\epsilon x't\over\hbar}-{1\over 2\sigma^2}
   (x'\mp {\epsilon t^2\over 2m})^2\right). \nonumber\\
\end{eqnarray}
Barring the exponential time-decay term, this represents a {\em pure state}
density matrix for a Gaussian wave-packet, centered at $x=\pm {\epsilon t^2
\over 2m}$, with a mean momentum $\bar{p}=\pm \epsilon t$. This has a nice
classical interpretation - with an acceleration $\epsilon /m$, after a
time $t$, the particle would have travelled a distance ${\epsilon t^2
\over 2m}$, and would have acquired a momentum $\epsilon t$. Thus one
sees that although there is fast decoherence, the classical dynamics
remains intact. So, the spin states are correlated with the wave-packets
moving in different directions.

One might be tempted to conclude from the preceding analysis that position
basis constitutes the pointer states. But before that we should also look at
what the spin-diagonal density matrix looks like in the momentum basis.
The spin diagonal elements of the reduced density matrix can be written,
in the momentum basis, as \cite{dk}
\begin{eqnarray}
\lefteqn{\rho_\pm(Q,q,t) =}\nonumber\\
&&  2\sqrt{\pi\over N(\tau)}\exp\left\{-{1\over N(\tau)}
\left(q\mp {\epsilon\over\hbar\gamma} (1-e^{-\tau})\right.\right.\nonumber\\
&& + \left. {i\hbar Q\over 2\sigma^2m\gamma}e^{-\tau}(1-e^{-\tau})
- {iDQ\over 4m\gamma^2\hbar}(1-e^{-\tau})^2 \right)^2 \nonumber\\
&&- \left({\hbar^2\over 4\sigma^2m^2\gamma^2}(1-e^{-\tau})^2 + {\sigma^2\over 4}\right.\nonumber\\
&&+\left. {D\over 2m^2\gamma^3}(2\tau-3+4e^{-\tau}-e^{-2\tau})\right)Q^2\nonumber\\
&&+\left.\left(\mp{i\epsilon\tau\over m\gamma^2} \pm {i\epsilon\over m\gamma^2}
(1-e^{-\tau})^2\right)Q\right\}, \label{pdensity}
\end{eqnarray}
where
\beq
N(\tau) = {D\over 2\hbar^2\gamma}(1-e^{-2\tau}) + {1\over\sigma^2}e^{-2\tau}.
\eeq

Again, this should be looked at in the meaningful limit $\gamma t \ll 1$.
In this limit, it takes the form,
\begin{eqnarray}
\rho_\pm(Q,q,t) &\approx& 2\sigma\sqrt{\pi}\exp\left(-{2\gamma k_BT(p-p')^2
t^3\over 3m\hbar^2}\right)   \nonumber\\
&&\times \exp\left(\mp{i\epsilon t^2p\over 2m\hbar}
-{\sigma^2\over 2\hbar^2}(p\mp\epsilon t)^2 \right)\nonumber\\ 
&&\times \exp\left(\pm{i\epsilon t^2p'\over 2m\hbar}
-{\sigma^2\over 2\hbar^2}(p'\mp\epsilon t)^2 \right),
\end{eqnarray}
where we have assumed that because of the exponential time decay due to
the first term, the
imaginary terms involving $Q$ in the exponent of (\ref{pdensity}), can 
be ignored.

Here, one can see that the momentum off-diagonal elements decay with time
in a way that is not strictly exponential because of the $t^3$ in the
exponent, instead of just $t$.  Nevertheless, one can extract a decay
time scale given by 
\beq
\tau_D = \left({3m\hbar^2\over 2\gamma k_BT(p-p')^2}\right)^{1/3}.\label{mytime}
\eeq
Barring the first term, the rest of the expression represents a {\em pure
state} density matrix for a momentum wave-packet, centered at $p=\pm\epsilon t$,
and with a mean position $\pm{\epsilon t^2\over 2m}$. This again has
a nice classical interpretation of a particle having travelled a distance
$\pm{\epsilon t^2\over 2m}$, and having acquired a momentum $\pm\epsilon t$,
because of an acceleration $\pm\epsilon/m$ due to the Stern-Gerlach field.

Thus the environment selects out Gaussian wave-packets approximately
localized both in position and momentum, as pointer states. So, the
``classicality" criterion, discussed in section II is satisfied here.
This result is in agreement with the earlier work by Eisert where it is shown
that decoherence to Gaussian states is generic for free particles coupled
to heat-bath \cite{eisert}. Here we see that even for the Stern-Gerlach setup, Gaussian
states emerge as pointer states, although with a centre that moves according
to classical dynamics. So,
the undamped Newtonian dynamics is recovered. Thus, the role of the
environment is only to localize spatially extended states. The Newtonain
dynamics, which is guaranteed by the Ehrenfest theorem, emerges naturally. 
With (approximately) well-defined position and momentum, one can now talk of
classical paths. So, because of decoherence, quantum particles acquire
trajectories.

Earlier works \cite{rg,dk} conclude that the pointer states for this problem are 
momentum states, and the decoherence time is given by (\ref{rgtime}).
Expression (\ref{rgtime}) cannot denote the true decoherence time, because it
depends inversely on the relaxation time of the particle. This would mean,
if the interaction with the environment, parametrized by $\gamma$ is decreased,
the decoherence should be faster. That cannot obviously be true. The catch
lies in the fact that in earlier works \cite{rg,dk}, the authors used the
limit $\gamma t \gg 1$, which takes one to the regime where there is not
just decoherence, but also dissipation. Decoherence happens very fast, in
the time within the limit $\gamma t \ll 1$. So, it is meaningful to use this
limit to study decoherence.

In the limit $\gamma t \gg 1$ dissipative effects overshadow the effect
of decoherence in the dynamics. This limit is physically meaningful in
studying the damped dynamics of quantum particles. The particle moving under
a potential, loses energy and ultimately thermalizes with the heat-bath,
ending in a meandering trajectory. For example, a quantum harmonic oscillator,
in this limit would be damped to the extent that it comes to its mean position
and executes thermal motion.

It should be pointed out that if pointer states were either
momentum states as claimed in \cite{rg,dk}, or position states as claimed
by Zurek \cite{zurek}, the density matrix would remain non-diagonal in the 
conjugate variable, and the classicality criteria won't be satisfied.
In our analysis, Gaussian states with moving centers, emerge as natural
pointer states. They look closer to classical states, and also 
yield {\em undamped} Newtonian dynamics.

\section{conclusions}

In this paper, we have clarified certain issues regarding the time-scales over
which environment-induced decoherence occurs, and the nature of emergent 
pointer states.
A model system, namely, the Stern-Gerlach setup, 
coupled to a quantum mechanical ``heat-bath" has been studied. The emergent
pointer states are Gaussian states which evolve following Newtonian
dynamics.


\begin{thebibliography}{99}
\bibitem{grw} G. C. Ghirardi, A. Rimini, and T. Weber, Phys. Rev. D {\bf 34}, 
470 (1986).

\bibitem{zeh} H.D. Zeh, Found. Phys. {\bf 1}, 69 (1970);
E. Joos and H.D. Zeh, Z. Phys. D {\bf 59}, 223 (1985).\\
For a recent review, see
{\em Decoherence and the appearance of a classical world in
quantum theory}, eds. Guilini et al. (Springer) (1996).

\bibitem{zurekrmp} {\em Decoherence, einselection, and the quantum origins
of the classical}, W. H. Zurek, Rev. Mod. Phys. {\bf 75}, 715-775 (2003).

\bibitem{feynman} R.P. Feynman and F.L. Vernon, Ann. Phys.(N.Y.)
{\bf 24}, 118 (1963).

\bibitem{dekker} H. Dekker, Phys. Rep. {\bf 80}, 1 (1981).

\bibitem{legget} A.O. Caldeira and A.J. Legget, Physica A {\bf 121},
587 (1983).

\bibitem{zurek} {\em Decoherence and the transition from quantum to
classical}, W. H. Zurek, Physics Today {\bf 44} (10), 36 (1991).

\bibitem{sieve} {\em Predictability sieve, pointer states, and the classicality of quantum trajectories}, D.A.R. Dalvit, J. Dziarmaga, W. H. Zurek, Phys. Rev. A {\bf 72}, 062101 (2005).

\bibitem{gogolin} {\em Environment-induced super selection without pointer states}, C. Gogolin, Phys. Rev. E {\bf 81}, 051127 (2010).

\bibitem{rg} {\em Quantum theory of a Stern-Gerlach system in contact
with a linearly dissipative environment}, S. Banerjee and R. Ghosh, 
Phys. Rev. A {\bf 62}, 042105 (2000).

\bibitem{dk} {\em Environment-induced decoherence I: The Stern-Gerlach 
measurement}, A. Venugopalan, D. Kumar, R. Ghosh, Physica {\bf 220}, 563
(1995).

\bibitem{anu1} {\em Preferred basis in a measurement process}, 
A. Venugopalan, Phys. Rev. A {\bf 50}, 2742 (1994).

\bibitem{anu} {\em Pointer states via decoherence in a quantum measurement}, 
A. Venugopalan, Phys. Rev. A {\bf 61}, 012102 (1999).

\bibitem{habib} 
W. H. Zurek, S. Habib and J. P. Paz, Phys. Rev. Lett. {\bf 70}, 1187 (1993).

\bibitem{paz} {\em Quantum limits of decoherence: environment induced 
super-selection of energy eigenstates}, J. P. Paz and W. H. Zurek, 
Phys. Rev. Lett. {\bf 82}, 5181 (1999).

\bibitem{eisert} {\em Exact Decoherence to Pointer States in Free Open Quantum Systems is Universal}, J. Eisert, Phys. Rev. Lett. {\bf 92}, 210401 (2004).

\end{thebibliography}
\end{document}